\documentclass{lipics-v2021}
\usepackage[utf8]{inputenc}
\usepackage{color} 

\hideLIPIcs \nolinenumbers

\usepackage{hyperref}
\usepackage{graphicx}

\DeclareMathOperator{\conv}{conv}
\newcommand{\CC}{\mathcal{C}}
\renewcommand{\SS}{\mathcal{S}}

\title{Convex Covering Using Collections of Convex Polygons and Set Cover}

\ccsdesc{Theory of computation~Computational geometry}
\keywords{Set cover, covering, polygons, convexity, heuristics, enumeration, simulated annealing, integer programming, computational geometry}
\supplementdetails[subcategory={Source Code}]{}{https://github.com/gfonsecabr/shadoks-CGSHOP2023}

 		\author{Guilherme D. da Fonseca}
 	{LIS, Aix-Marseille Université}
 	{guilherme.fonseca@lis-lab.fr}
 	{https://orcid.org/0000-0002-9807-028X}
 	{}
 	
 	\authorrunning{Guilherme D. da Fonseca}

\funding{Work supported by the French ANR PRC grant ADDS (ANR-19-CE48-0005).}

\Copyright{Guilherme D. da Fonseca}

\acknowledgements{We would like to thank the Challenge organizers and other competitors for their time, feedback, and making this whole event possible. We would also like to thank the second year undergraduate students Rayis Berkat, Ian Bertin, Ulysse Holzinger, and Julien Jamme for coding a solution visualisation tool that allows you to view all our best solutions in \url{https://pageperso.lis-lab.fr/guilherme.fonseca/cgshop23view/}.
}

\begin{document}

\maketitle

\begin{abstract}
In the convex covering problem, we are given a convex polygon with holes $P$ and the goal is to cover $P$ using a small number of convex polygons that lie inside $P$. In this paper, we solve the problem using the following strategy. We find a big collection of large (often maximal) convex polygons inside $P$ and then solve several set cover problems to find a small subset of the collection that covers the whole polygon. The quality of our heuristics is confirmed by winning the second place in the CG:SHOP 2023 Challenge.
\end{abstract}

\section{Introduction} \label{s:intro}

\begin{figure}[b]
 \centering
 \includegraphics[height=4.3cm]{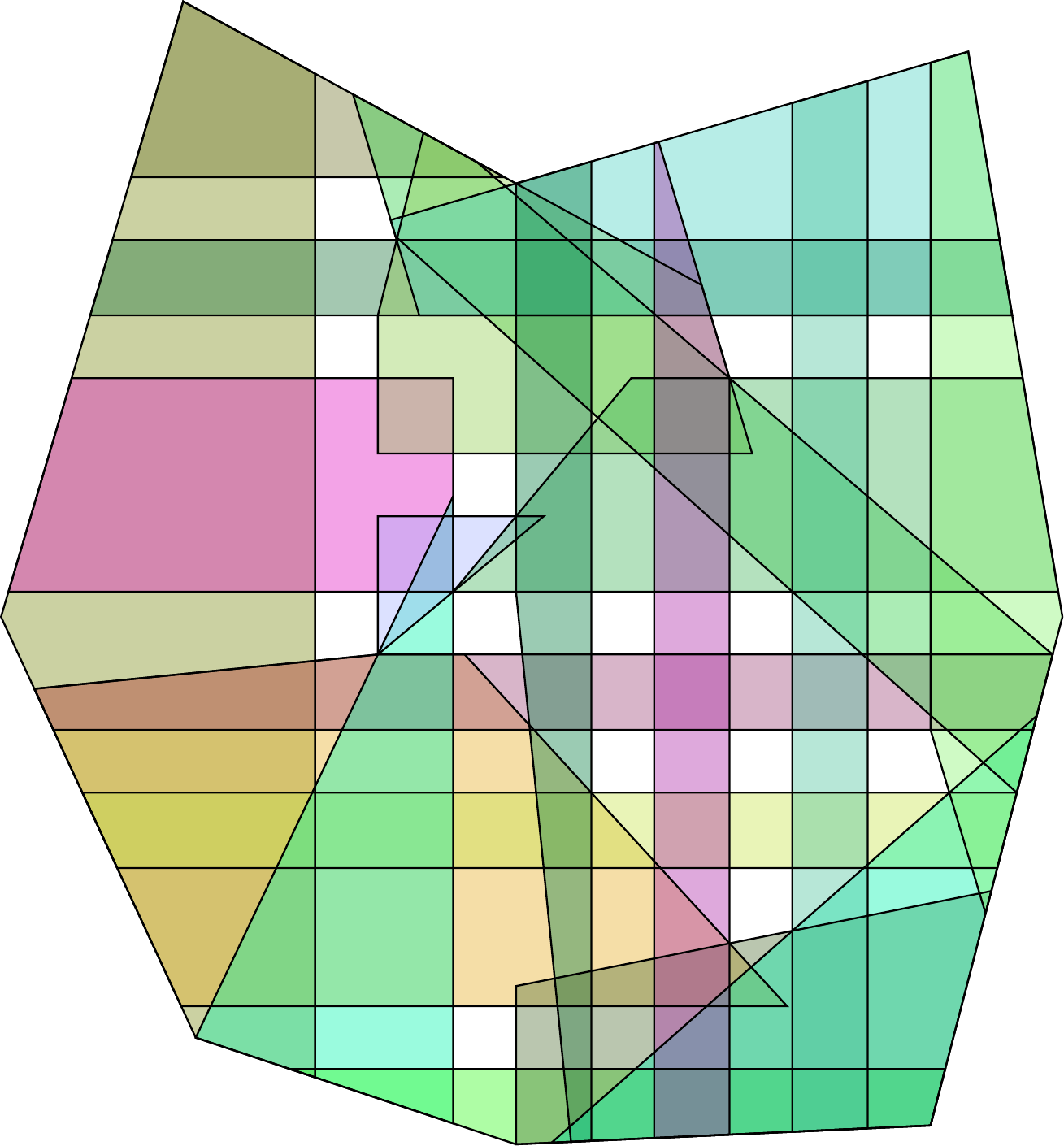} \hspace{1em}
 \includegraphics[height=4.3cm]{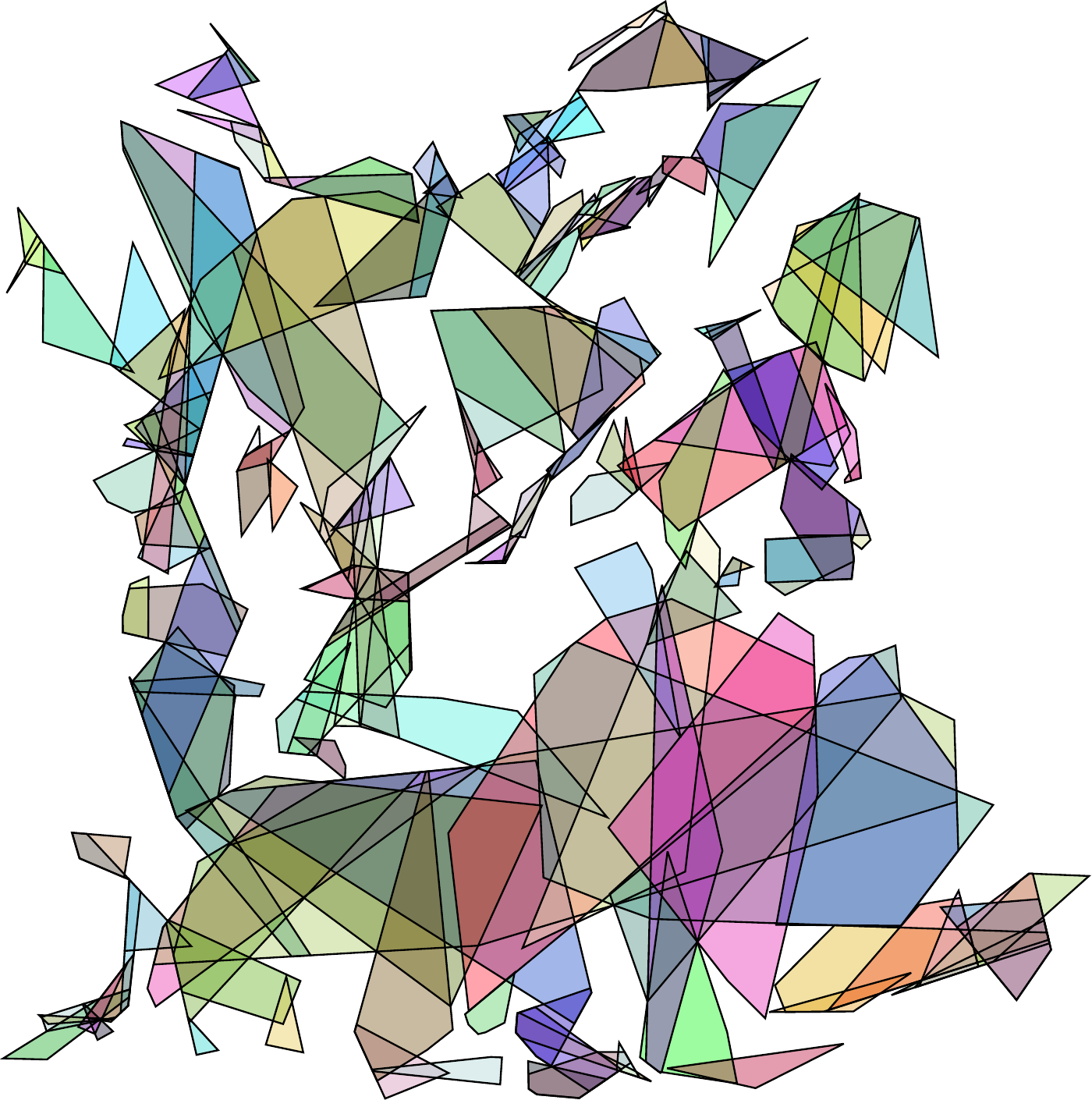} \hspace{1em}
 \includegraphics[height=4.3cm]{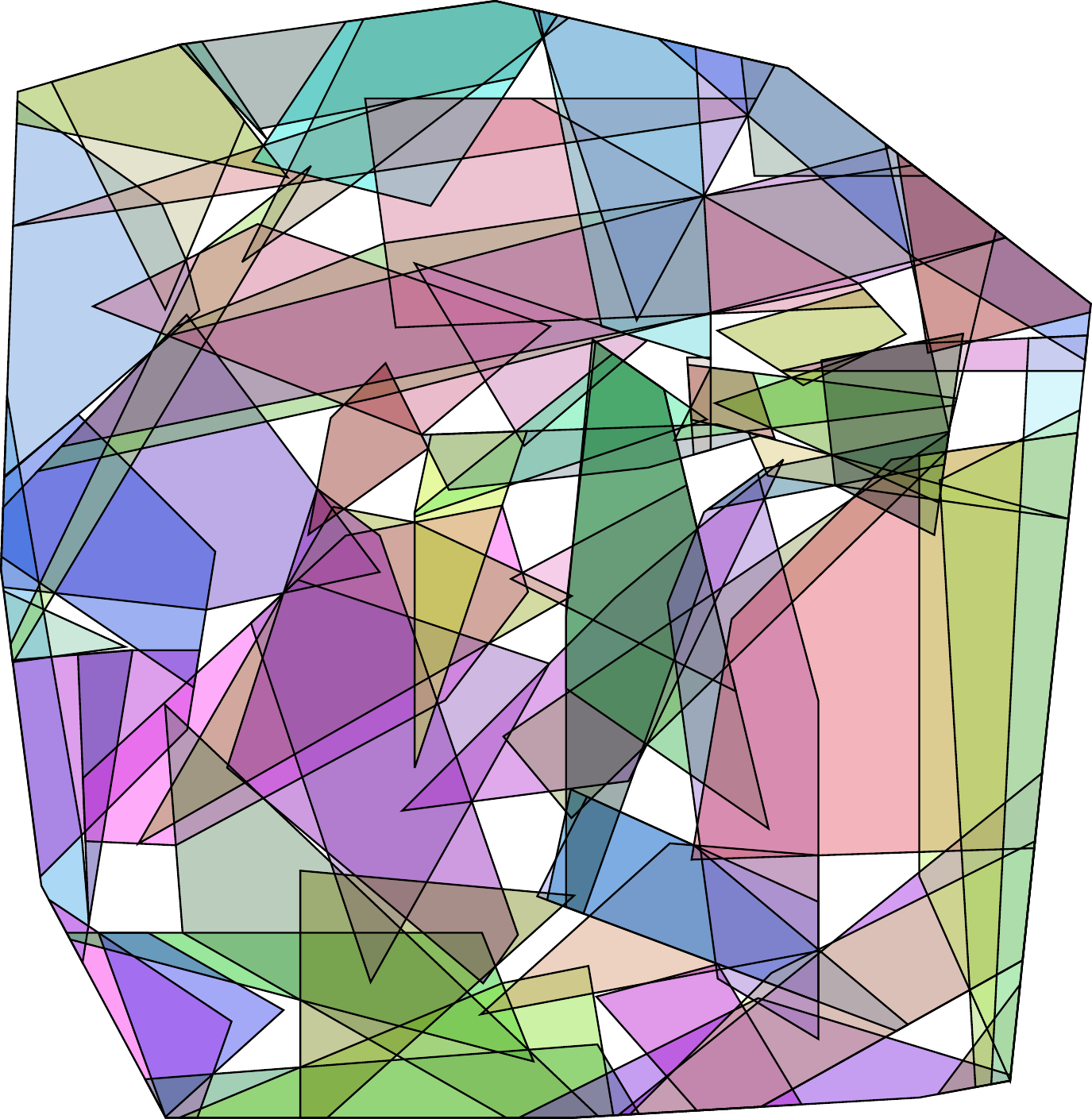}
 \caption{Solutions to \texttt{maze\_79\_50\_5\_5}, \texttt{fpg-poly\_400\_h2}, and \texttt{cheese163} instances. The instances have $64$, $462$, and $163$ vertices and the solutions have $23$, $172$, and $68$ polygons, respectively.}
 \label{f:sols}
\end{figure}

CG:SHOP Challenge is an annual geometric optimization challenge. The fifth edition in 2023 considers the problem of covering a polygon with holes $P$ using a small set of convex polygons $\SS$ that lie inside $P$. 
In total, $206$ polygons have been given as instances, ranging from $24$ to $109{,}360$ vertices. The instances are of several different types, including orthogonal polygons and polygons with many small holes. The team Shadoks won second place with the best solution (among the $22$ participating teams) to $128$ instances. More details about the Challenge and this year's problem are available in the organizers' survey paper~\cite{survey}.

Our general strategy consists of two distinct phases. First, we produce a large \emph{collection} $\CC$ of large convex polygons inside $P$. Second, we find a small subset $\SS \subseteq \CC$ that covers $P$, which is returned as the solution. Figure~\ref{f:sols} shows three small solutions and we can observe that most convex polygons are maximal and often much larger than necessary. Our approach is different from that of the winning team DIKU (AMW), that uses clique cover of a different graph~\cite{diku}.

To construct the collection $\CC$ in phase $1$, we use either a modified version of the Bron-Kerbosch algorithm~\cite{BrKe73} or a randomized bloating procedure starting from a constrained Delaunay triangulation (Section~\ref{s:collections}). We then construct and solve a set cover problem in phase~$2$. To solve the set cover problem, we use integer programming and simulated annealing. The key element for the efficiency of phase $2$ is to iteratively generate constraints as detailed in Section~\ref{s:setcover}. Generally speaking, the initial constraints ensure that all input vertices are covered and supplementary constraints ensure that a point in each uncovered area is covered in the following iteration.

\begin{figure}
  \centering
  \begin{minipage}{8cm}
    \raisebox{1.2cm}{$162$}
    \includegraphics[width=2.7cm]{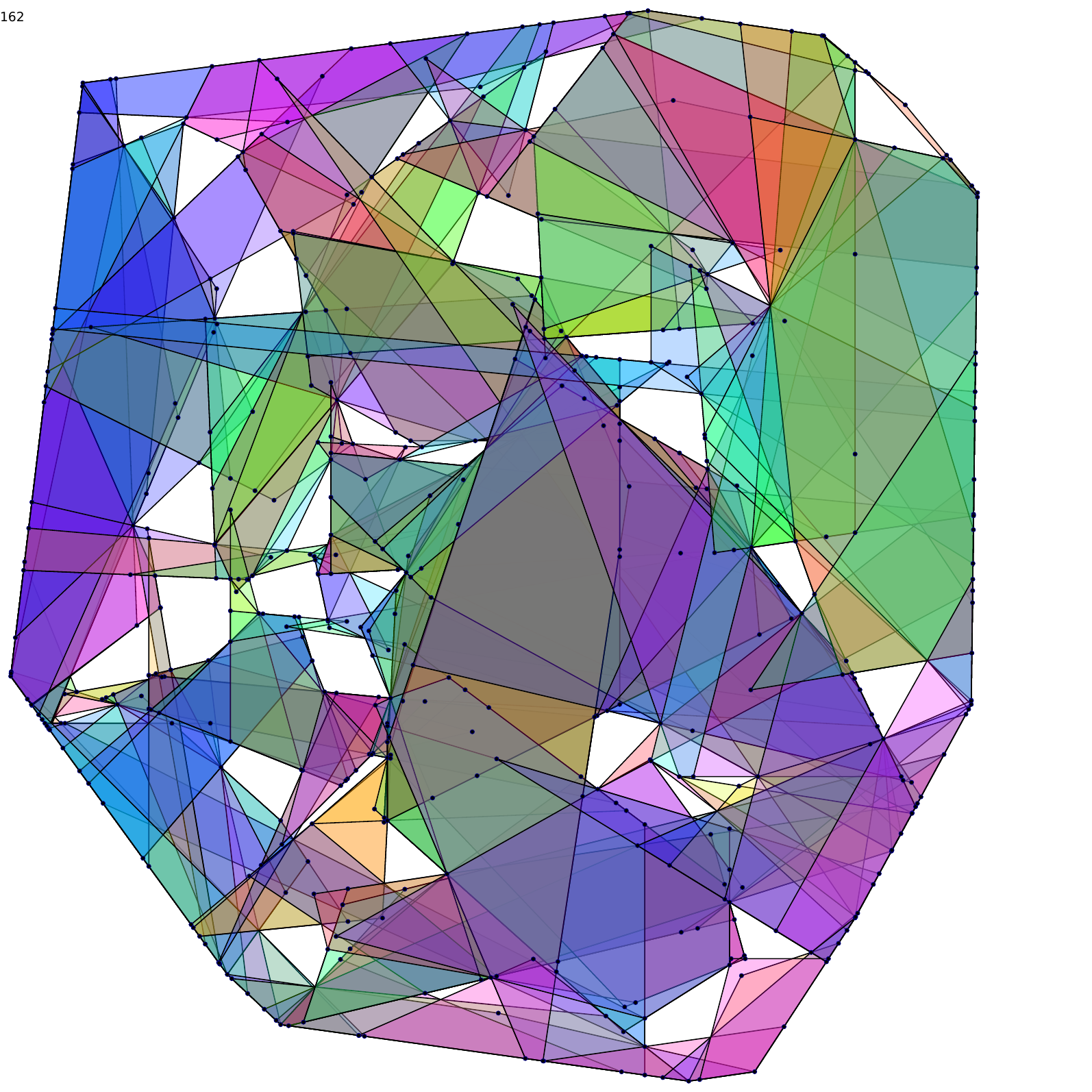}
    \raisebox{1.2cm}{$\longrightarrow\; 74$}
    \includegraphics[width=2.7cm]{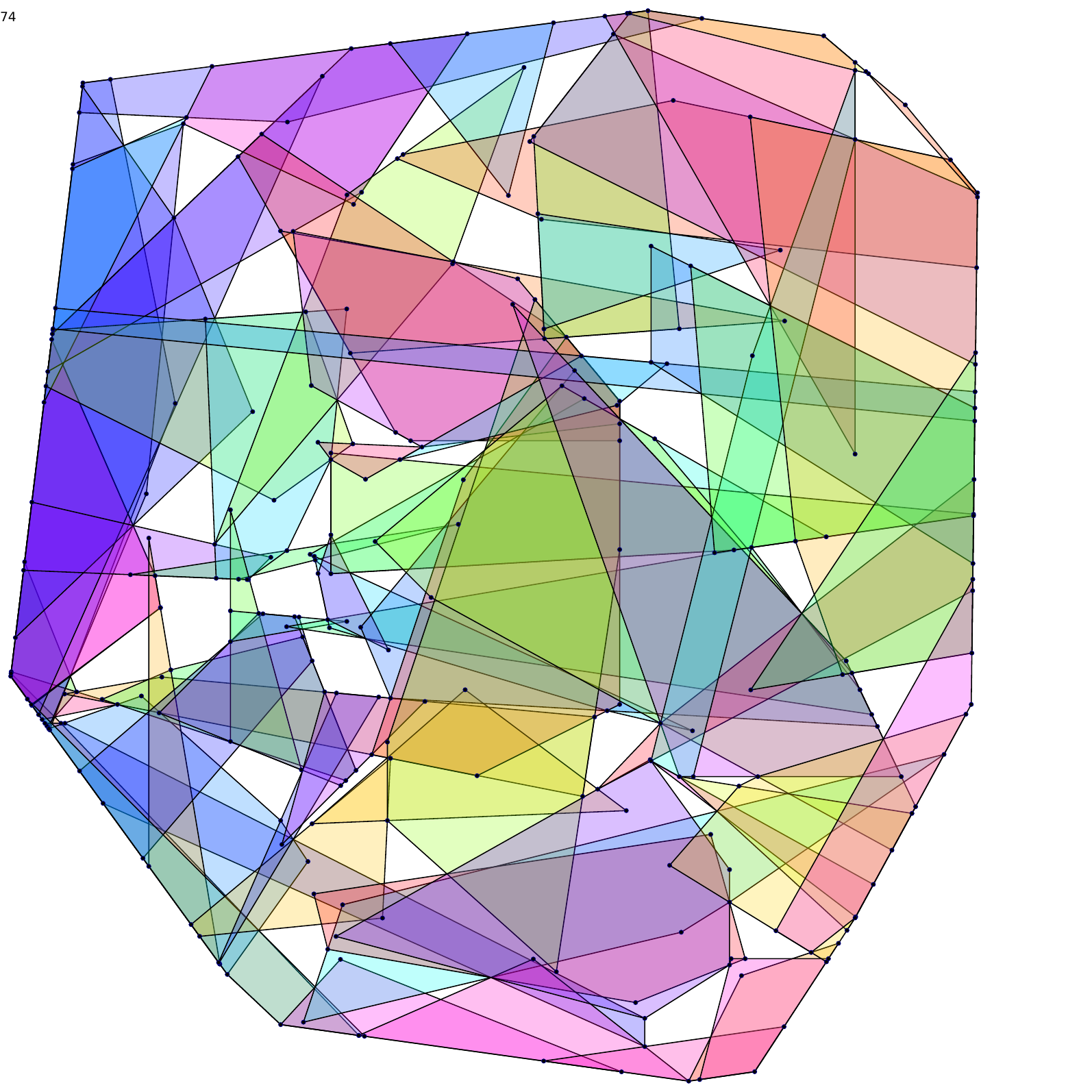}
    \raisebox{1.2cm}{$\searrow$}
    \\ \medskip \\
    \raisebox{1.2cm}{$160$}
    \includegraphics[width=2.7cm]{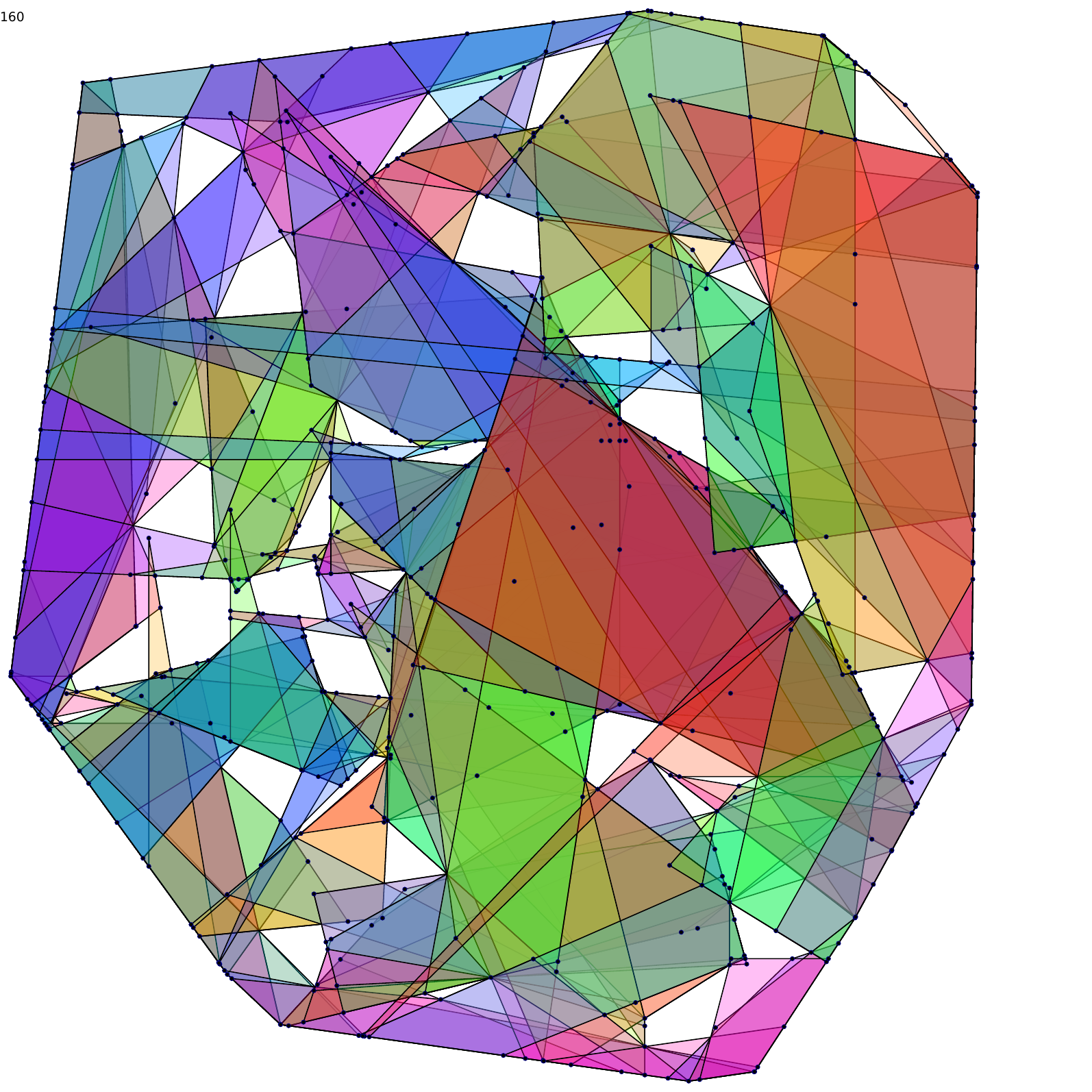}
    \raisebox{1.2cm}{$\longrightarrow\; 72$}
    \includegraphics[width=2.7cm]{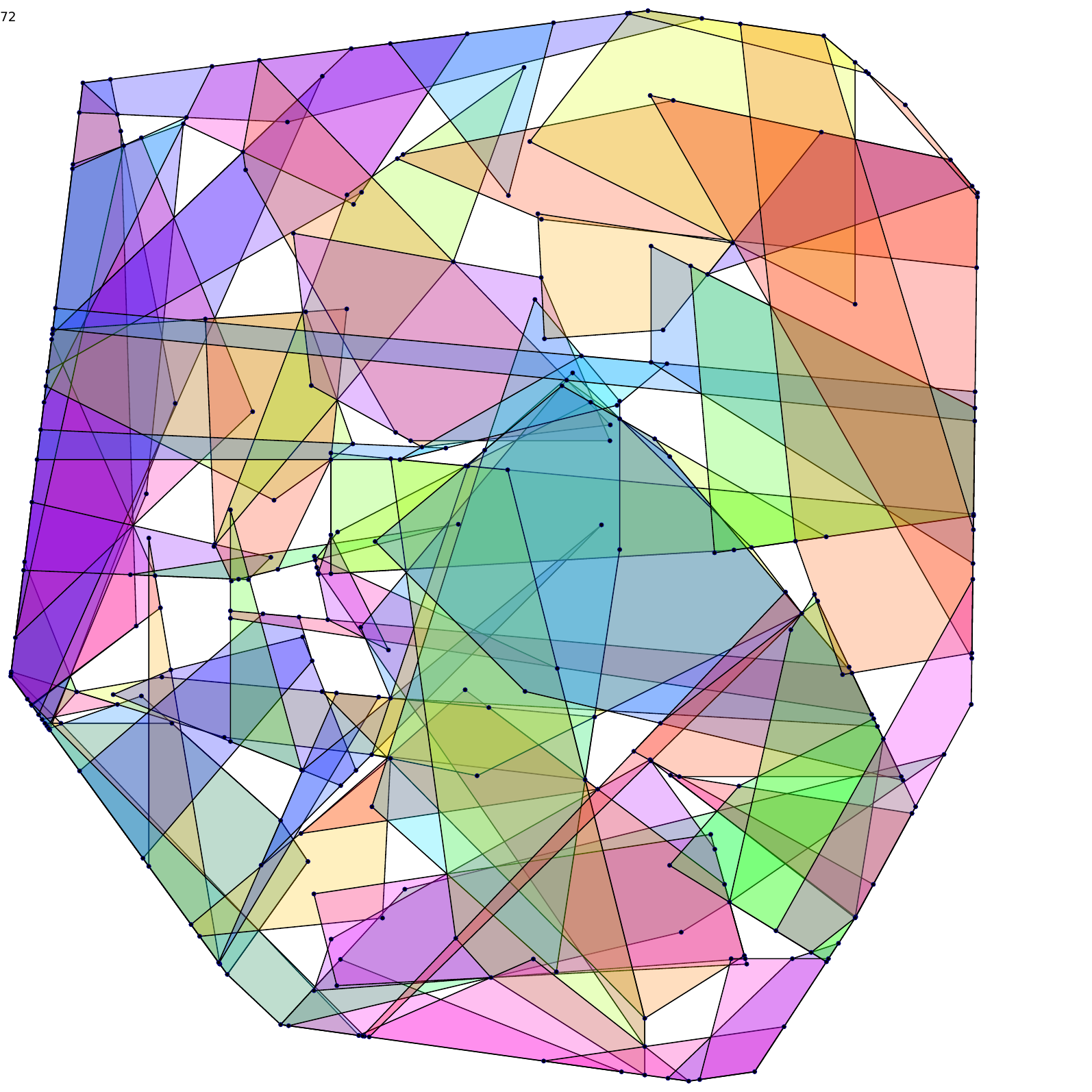}
    \raisebox{1.2cm}{$\nearrow$}
  \end{minipage}\begin{minipage}{3.5cm}
    $70$
    \raisebox{-0.5\height}{\includegraphics[width=2.7cm]{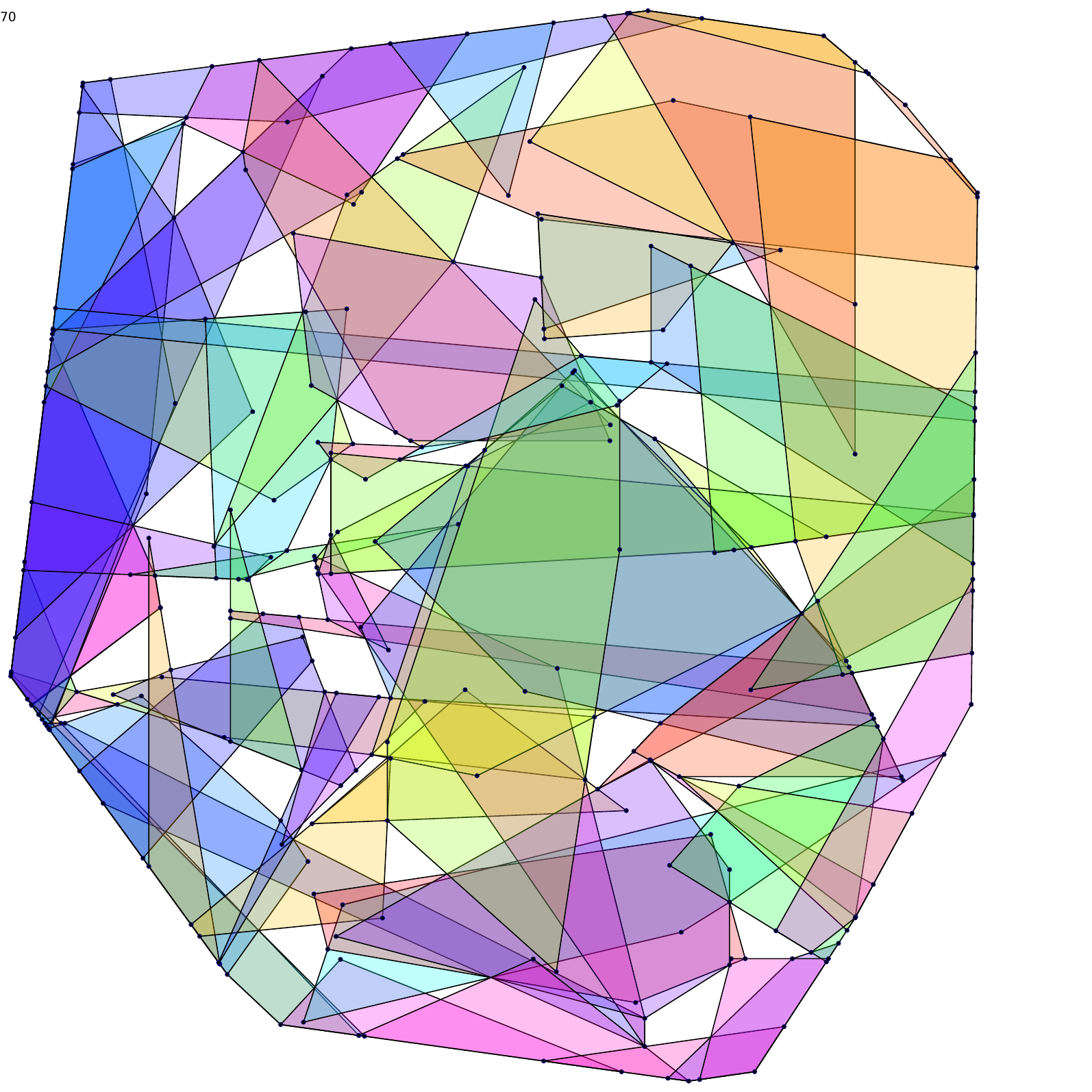}}
  \end{minipage}
  \caption{Two collections with $162$ and $160$ large convex polygons are fed into the set cover solver that respectively produces subsets with $74$ and $72$ that cover the input polygon. The two subsets are united into a new collection and the set cover solver finally produces a solution with only $70$ convex polygons for instance \texttt{ccheese142}.}
  \label{f:mergeexample}
\end{figure}

In fact, to obtain our best solutions, we repeat phase $2$ using the union of the solutions from independent runs of the first two phases as the collection $\CC$. An illustrative example is presented in Figure~\ref{f:mergeexample}. The results we obtained are discussed in Section~\ref{s:results}.

\section{Collections} \label{s:collections}

We  now describe phase $1$ of our strategy: building a collection. Throughout, the instance is a polygon with holes $P$ with vertex set $V$. Formally speaking, a \emph{collection} $\CC$ is defined exactly as a \emph{solution} $\SS$: a finite set of convex polygons whose union is $P$. However, while we want a solution $\SS$ to have as few elements as possible, the most important aspect of a collection $\CC$ is that it contains a solution $\SS \subseteq \CC$ with few elements. Ideally, $|\CC|$ is also not too big so the phase 2 solver is not overloaded, but the size of $\CC$ is of secondary importance.

Before we describe two methods to build collections, we present a definition.
Given a set of points $S$, a convex polygon $C \subseteq P$ is \emph{$S$-maximal} if the vertices of $C$ are in $S$ and there exists no point $s \in S$ with $\conv(C \cup \{s\}) \subseteq P$. Next, we show how to build a collection with all $S$-maximal convex polygons.

\subparagraph*{Bron-Kerbosch}

The Bron-Kerbosch algorithm~\cite{BrKe73} is a classic algorithm to enumerate all maximal cliques in a graph (in our case a visibility graph where a pair of vertices defines an edge when the corresponding line segment is contained in the polygon) with good practical performance~\cite{ELS13,Koc01}. The algorithm recursively keeps the following three sets and its pseudo-code is presented in Listing~\ref{l:bk}.
\begin{itemize}
 \item[$\mathtt{R}$:] Vertices in the current maximal clique. Initially, $\mathtt{R}=\emptyset$.
 \item[$\mathtt{S}$:] Vertices that may be added to the current maximal clique (these must be adjacent to all vertices in $\mathtt{R}$). Initially, $\mathtt{S} = S$.
 \item[$\mathtt{X}$:] Vertices that may not be added to the current maximal clique because otherwise the same clique would be reported multiple times. Initially, $\mathtt{X}=\emptyset$.
\end{itemize}

\begin{lstlisting}[caption={Bron-Kerbosch algorithm to enumerate the maximal cliques in a graph. The neighborhood of a vertex \texttt{v} is denoted \texttt{N(v)}.},label=l:bk,float=t,abovecaptionskip=-\medskipamount,escapeinside={(*}{*)}]
BronKerbosch(R, S, X):
  if S and X are both empty:
    report R as a maximal clique
    return

  for each vertex v in S:
    S' (*$\gets$*) S (*$\cap$*) N(v)
    X' (*$\gets$*) X (*$\cap$*) N(v)

    BronKerbosch(R (*$\cup$*) {v}, S', X')

    S (*$\gets$*) S \ {v}
    X (*$\gets$*) X (*$\cup$*) {v}
\end{lstlisting}

Next, we show how to adapt this algorithm to enumerate all $S$-maximal convex polygons. As there are infinitely many $P$-maximal polygons, it is essential to restrict our set $S$. Different sets $S$ will be described later in this section.

\begin{figure}[p]
 \centering
 \includegraphics[scale=.42]{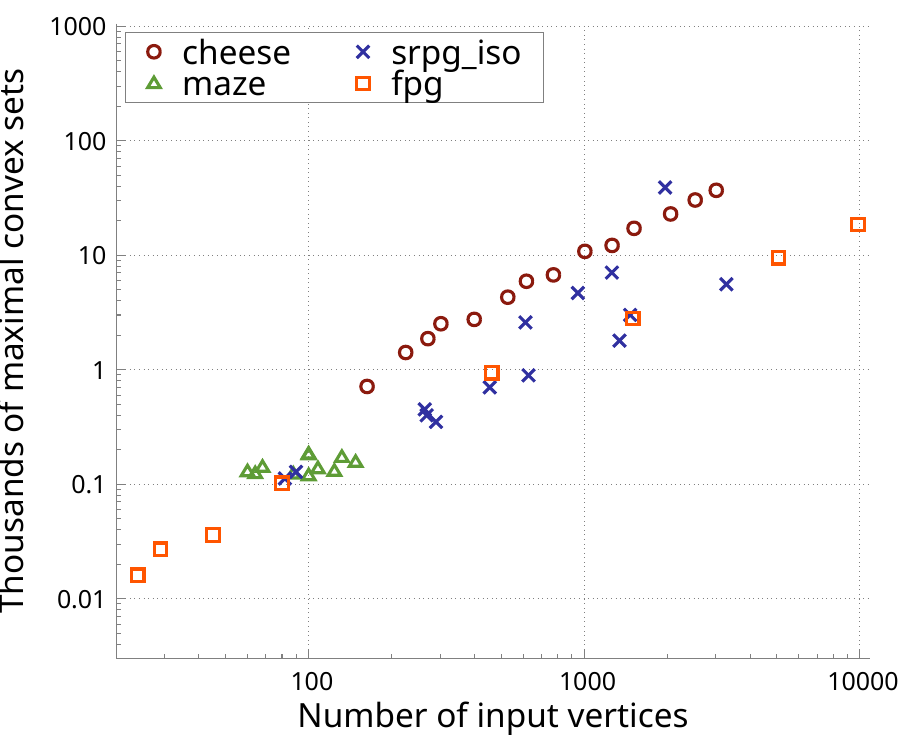} \hspace{2em}
 \includegraphics[scale=.42]{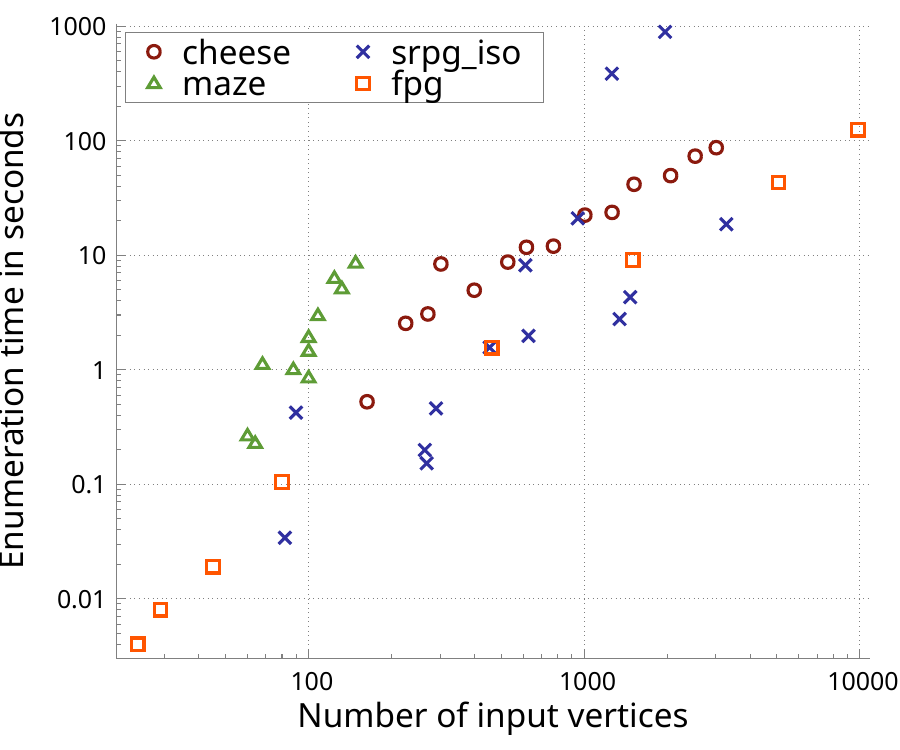}
 \caption{Number of $V$-maximal convex polygons and the time to enumerate them.}
 \label{f:max}
\end{figure}

If the polygon $P$ has no holes and $S$, then there is a bijection between the maximal cliques in the visibility graph of $S$ with respect to $P$ and the $S$-maximal convex polygons. While this statement is no longer true when $P$ has holes, we can adapt the Bron-Kerbosch algorithm to enumerate all $S$-maximal convex polygons as shown in Listing~\ref{l:mbk}, that takes the polygon holes into account.
Figure~\ref{f:max} shows how the number of $V$-maximal convex polygons grows for different instances and that we can compute all $V$-maximal convex polygons quickly for instances with around 10 thousand vertices.

\begin{figure}[ht]
 \centering
 \includegraphics[scale=.85]{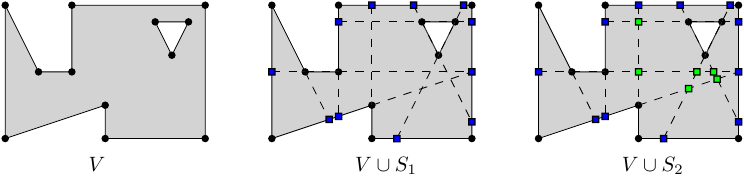}
 \caption{Definitions of $V$, $S_1$, and $S_2$.}
 \label{f:sets}
\end{figure}

\begin{lstlisting}[caption={Modified Bron-Kerbosch algorithm to enumerate the $S$-maximal convex polygons.},label=l:mbk,float=p,abovecaptionskip=-\medskipamount,escapeinside={(*}{*)}]
BronKerbosch(R, S, X):
  if S and X are both empty:
    report conv(R) as a maximal convex polygon
    return
  if X (*$\cap$*) conv(R) (*$\neq$*) {}:
    return // Only for improved performance

  for each vertex v in S:
    S' (*$\gets$*) X' (*$\gets$*) {}
    for each vertex u in S:
      if v (*$\neq$*) u and conv(R (*$\cup$*) {u,v}) (*$\subseteq$*) P:
        S' (*$\gets$*) S' (*$\cup$*) {u}

    for each vertex u in X:
      if v (*$\neq$*) u and conv(R (*$\cup$*) {u,v}) (*$\subseteq$*) P:
        X' (*$\gets$*) X' (*$\cup$*) {u}

    BronKerbosch(R (*$\cup$*) {v}, S', X')

    S (*$\gets$*) S \ {v}
    X (*$\gets$*) X (*$\cup$*) {v}
\end{lstlisting}

\begin{figure}[p]
 \centering
 \includegraphics[scale=.42]{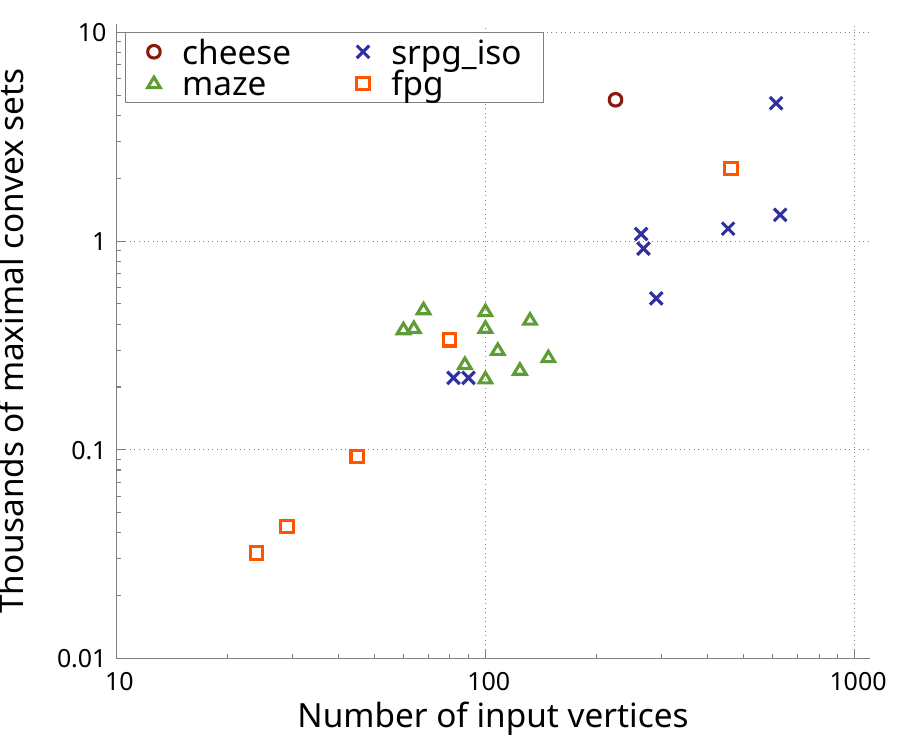} \hspace{2em}
 \includegraphics[scale=.42]{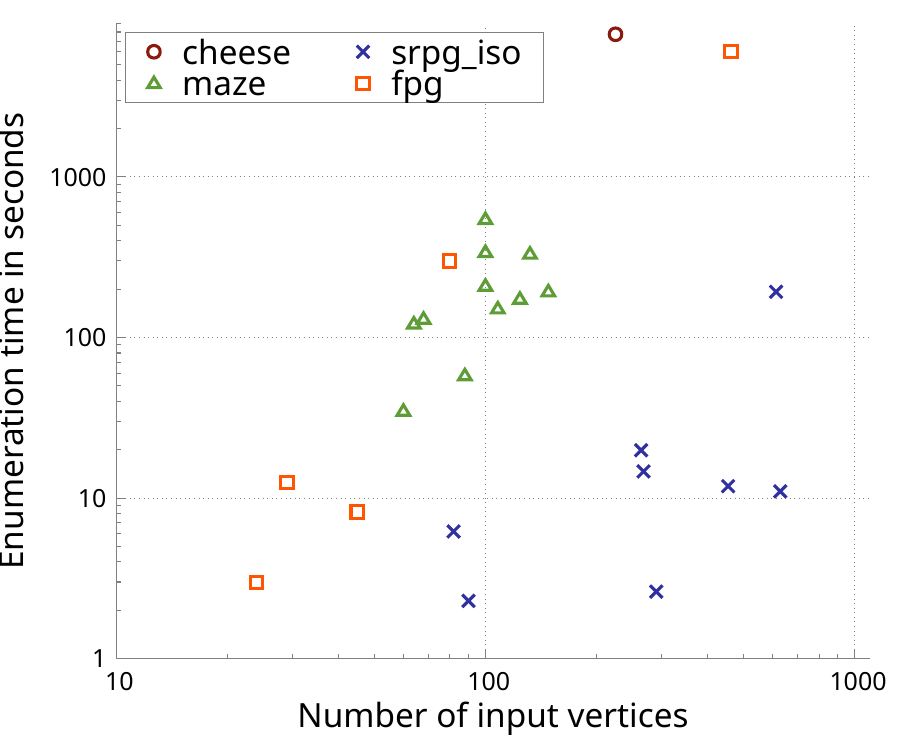}
 \caption{Number of $(V \cup S_1)$-maximal convex polygons and the time to enumerate them.}
 \label{f:maxp}
\end{figure}

Let $S_1$ be the set of the endpoints of the largest segments inside $P$ that contain each edge of $P$ (Figure~\ref{f:sets}). It is easy to see that $|S_1| \leq 2 |V|$. However, as shown in Figure~\ref{f:maxp}, we are only able to compute all $(V \cup S_1)$-maximal convex polygons for instances with less than one thousand vertices. It is possible that a modified version of the Bron-Kerbosch algorithm gives better results, either by using a pivot or choosing a particular order for the points, but we have not succeeded in obtaining significant improvements.
Another natural set of points is the set $S_2 \supseteq S_1$ defined as the intersection points (inside $P$) of the lines containing the edges of $P$ (Figure~\ref{f:sets}). The set $S_2$ may however have size roughly $|V|^2$. Hence, computing all  $(V \cup S_2)$-maximal convex polygons is only feasible for very small instances.

\subparagraph*{Random Bloating}

As a $V$-maximal convex polygon $C$ is generally not $P$-maximal, we also grow $C$ with an operation we call bloating. Given a convex polygon $C$ and a set of points $S$, we construct an \emph{$S$-bloated} convex polygon $C'$ by iteratively trying to add a random point from $S$ to $C$ and taking the convex hull, verifying at each step that $C'$ lies inside the instance polygon $P$. There are two sets of points that may compose $S$. First, $S_1(C)$ is the set of endpoints of the largest segment in $P$ that contains each edge of $C$. Second, $S_2(C)$ is the union of $S_1(C)$ and the intersection points of the lines containing the edges of $C$, if the points are inside $P$. Notice that $|S_1(C)| = O(|C|)$, but $|S_2(C)| = O(|C|^2)$.

To start the bloating operation, we need a convex polygon $C$. One approach is to use the $V$-maximal convex polygons produced by Bron-Kerbosch. A much faster approach for large instances is to use a constrained Delaunay triangulation of the instance polygon. In this case, we start by $V$-bloating the triangles into a convex polygon $C$, and then possibly $S_1(C)$-bloating or $S_2(C)$-bloating the polygon $C$.
Since the procedure is randomized, we can \emph{replicate} the triangles multiple times to obtain larger collections of large convex polygons.

\section{Set Cover} \label{s:setcover}

We  now describe phase $2$ of our strategy. The input of phase $2$ is a collection $\CC$ of convex polygons that covers the instance polygon $P$, and the goal is to find a small subset of $\CC$ that still covers $P$.

\begin{figure}[ht]
 \centering
 \includegraphics[width=6cm]{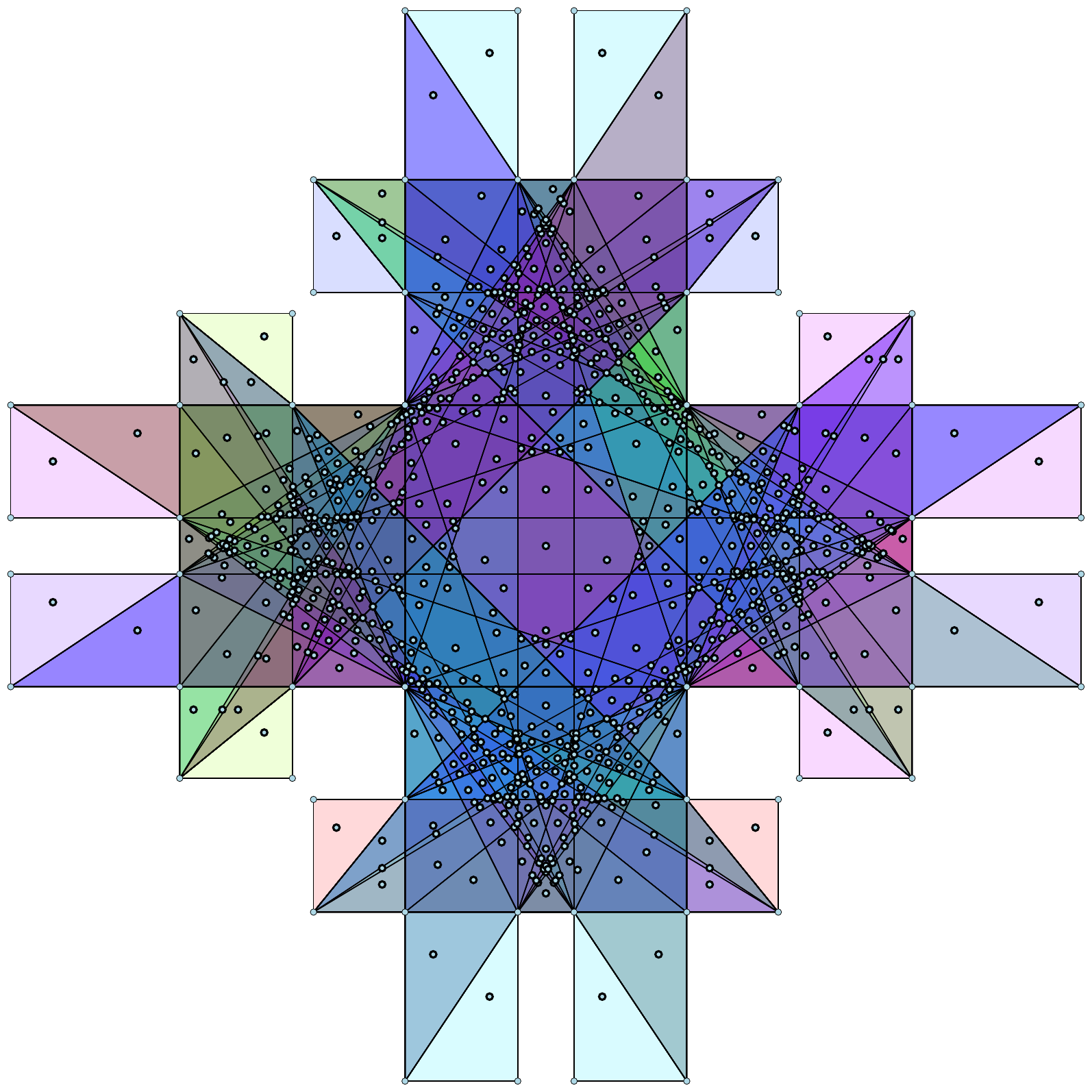} \hspace{1cm}
 \includegraphics[width=6cm]{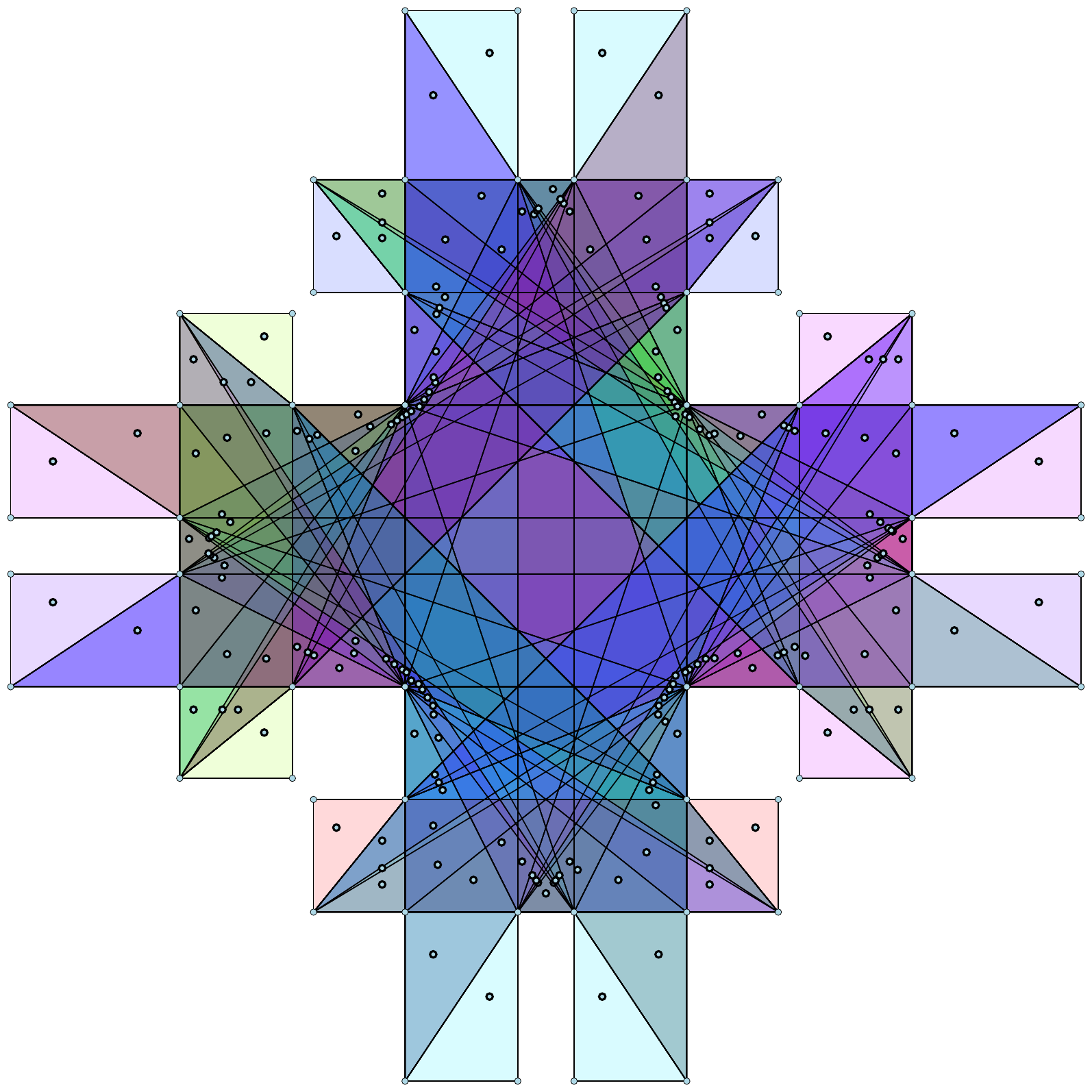}\\
 (a) \hspace{6.5cm} (b)
 \caption{All the 82 $V$-maximal convex polygons for the \texttt{socg\_fixed60} instance with (a) the 1009 arrangement witnesses and (b) the 200 vertex witnesses.}
 \label{f:witnesses}
\end{figure}

In contrast to the classic set cover problem, in our case $P$ is an infinite set of points. Nevertheless, it is easy to create a finite set of \emph{witnesses} $W$, that satisfy that $W$ is covered by a subset $\SS$ of $\CC$ if and only if $P$ is. To do that, we place a point inside each region (excluding holes) of the arrangement of line segments defining the boundaries of the polygons in $\CC$ and $P$ (Figure~\ref{f:witnesses}(a)). The size of such set $W$ is however very large in practice and potentially quadratic in the number of segments of $\CC$.

Producing small sets of witnesses has been studied in the context of art gallery problems~\cite{CJKM06}. However, we do not know if small sets of witnesses exist for our problem. Hence, we use a loose definition of witness as any finite set of points $W \subset P$. In practice, we want $W$ to be such that if $W$ is covered, then most of $P$ is covered. Next, we show how to build such set of witnesses and afterwards we describe how we solve the finite set cover problem.

\subparagraph*{Witnesses}

A set of witnesses $W$ that gave very good results, which we call \emph{vertex witnesses}, consists of one witness inside each cell of the arrangement that contains a vertex of the instance polygon $P$, as shown in Figure~\ref{f:witnesses}(b). This set guarantees that if $W$ is covered, then all points that are arbitrarily close to the \emph{vertices} of $P$ are covered. However, trivially computing $W$ requires building the arrangement of the collection $\CC$, which is too slow and memory consuming for large $\CC$.

A set of witnesses $W$ that also gives excellent results and is much faster to compute is called \emph{quick vertex witnesses}.
For each vertex $v$ of $P$, we consider all edges in $\CC$ and also $P$ that are adjacent to $v$. We order these edges around $v$ starting and ending with the edges of $P$. For each pair of consecutive edges, we add a point $w$ to $W$ that is between the two consecutive edges and infinitely close to $v$. Notice that the number of vertex witnesses is linear in the number of edges of $\CC$ and it can also be built in near linear time, avoiding the construction of the whole arrangement of $\CC$. If $P$ has no colinear points, then the quick vertex witnesses give the same vertex coverage guarantee as the vertex witnesses. We represent points that are arbitrarily close to $v$ implicitly as a point and a direction.

\begin{figure}[t]
 \centering
 \includegraphics[width=4cm]{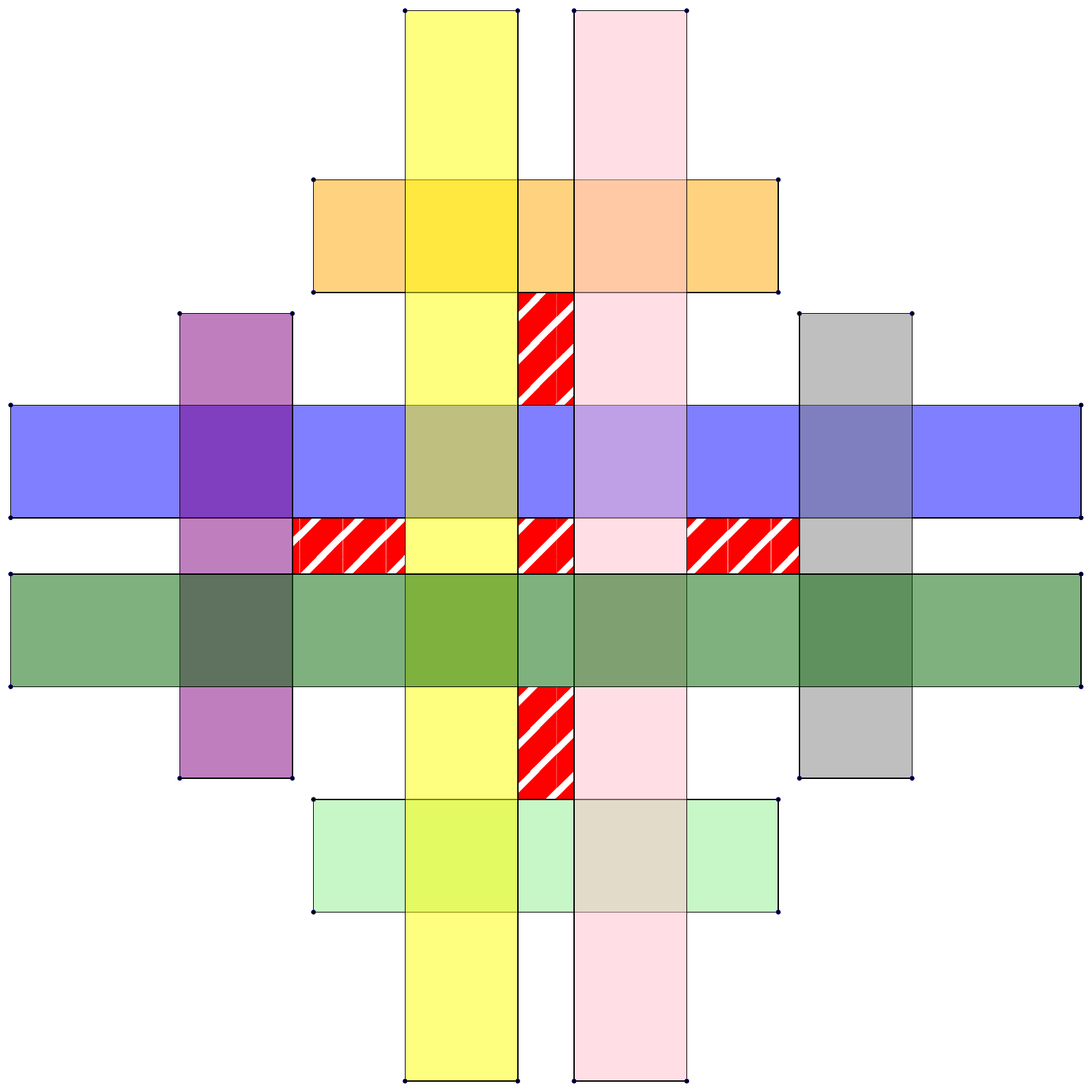} \hspace{1cm}
 \includegraphics[width=4cm]{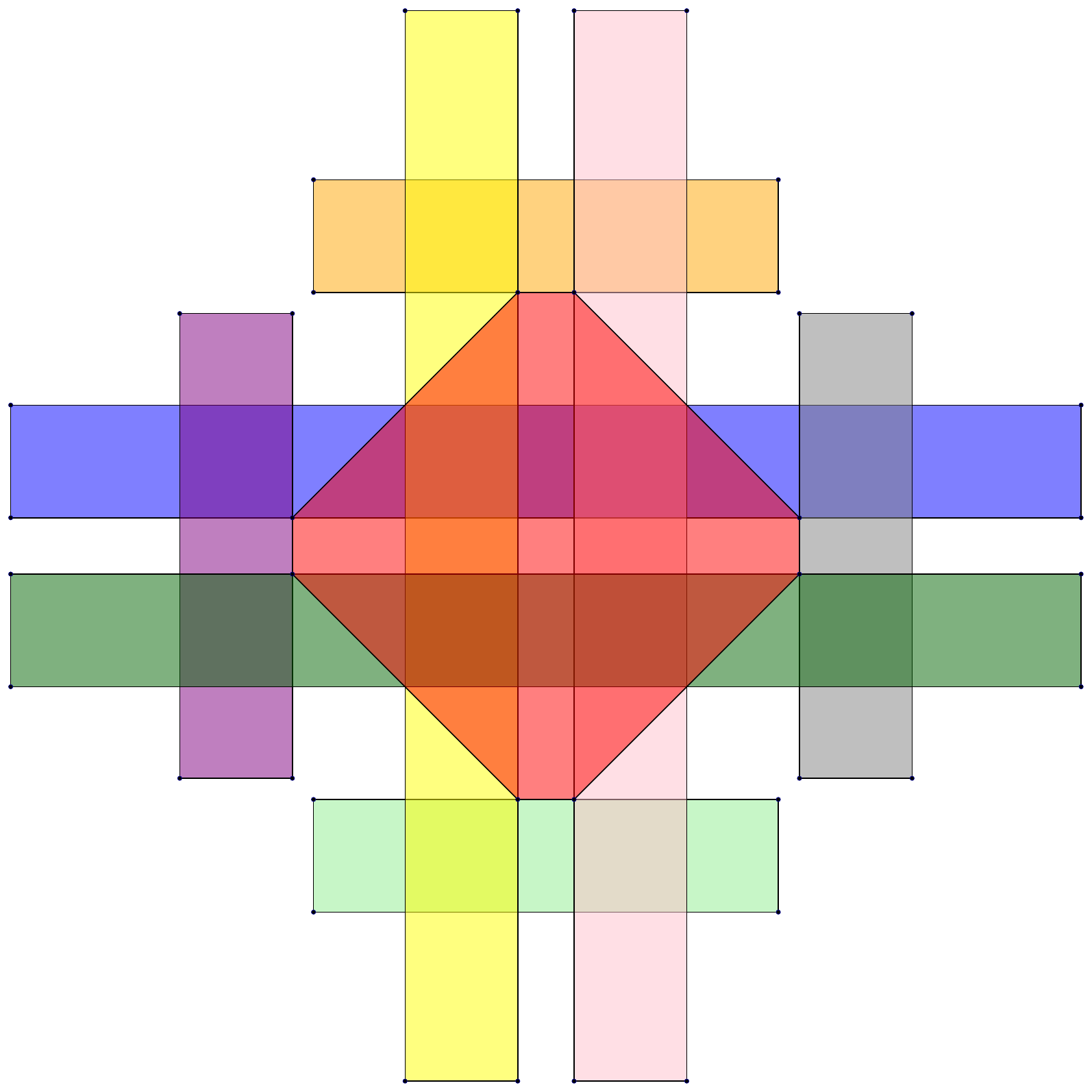}\\
 (a) \hspace{4.5cm} (b)
 \caption{(a) A solution that covers all vertex witnesses of the \texttt{socg\_fixed60} instance but not the whole polygon. Uncovered regions are marked in striped red. (b) The optimal solution obtained from the previous one by merging the uncovered regions.}
 \label{f:socg}
\end{figure}

Given a set $\SS'$ of convex polygons that cover $W$, there are two natural options to produce a valid solution $\SS$. The first option is to make $\SS = \SS' \cup \mathcal{R}$ for a set $\mathcal{R}$ built as follows. The \emph{uncovered region} $P \setminus \cup_{C \in \SS'}C$ consists of a set $\mathcal{U}$ of disjoint polygons, possibly with holes (Figure~\ref{f:socg}(a)). However, most of the time the polygons in $\mathcal{U}$ are in fact convex. For each polygon $U \in \mathcal{U}$, if $U$ is convex, then we add $U$ to $\mathcal{R}$. Otherwise, we triangulate $U$ and add the triangles to $\mathcal{R}$. Furthermore, we can greedily merge convex polygons in $\mathcal{R}$ to reduce their number, as long as the convex hull of the union remains inside $P$, which works very well for the SoCG logo solution shown in Figure~\ref{f:socg}(b).

\begin{figure}[ht]
 \centering
 \includegraphics[scale=.42]{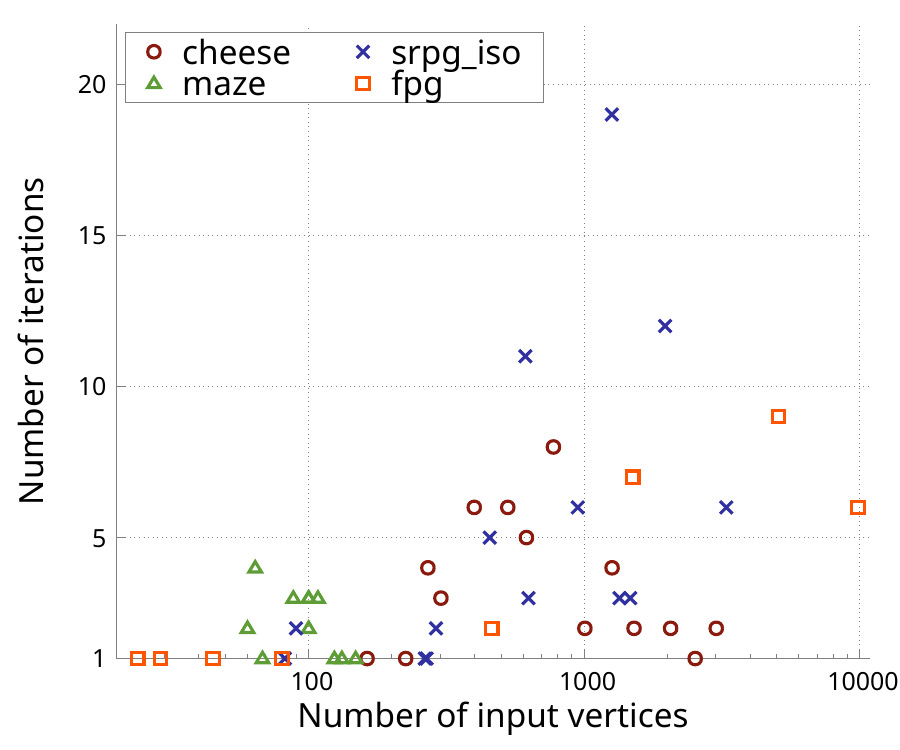} \hspace{2em}
 \includegraphics[scale=.42]{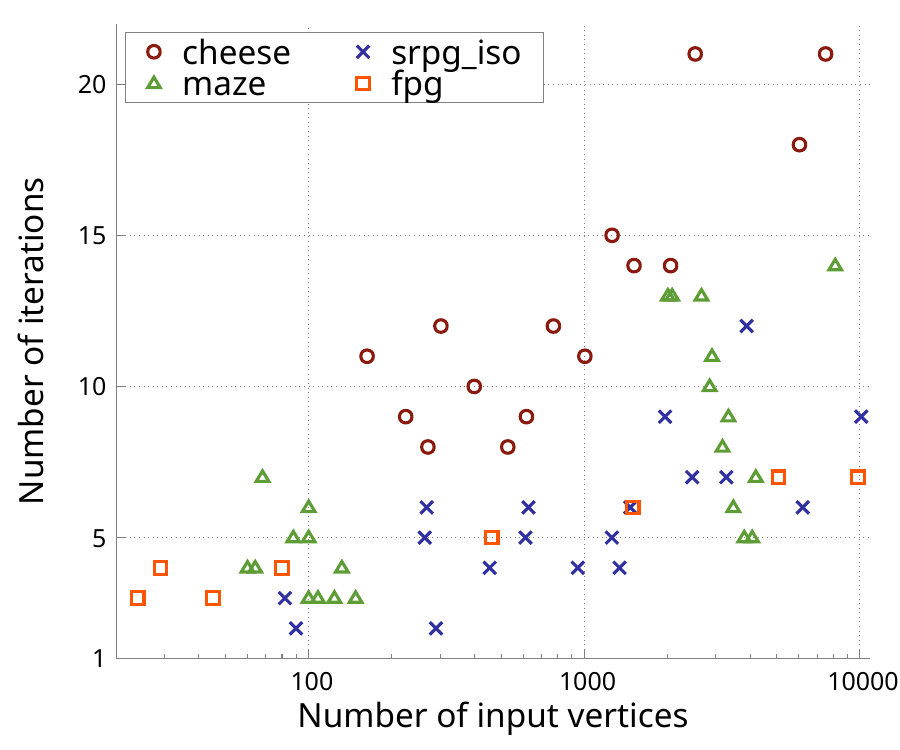}\\
  (a) \hspace{6cm} (b)
 \caption{Number of iterations to find a valid solution starting from quick vertex witnesses using IP as the solver, setting $\CC$ as (a) all $V$-maximal convex polygons and (b) $2$ times triangulation $(V \cup S_2(C))$-bloated convex polygons. Some large instances shown in (b) do not appear in (a) because enumearting all $V$-maximal convex polygons would take too long.}
 \label{f:vwit}
\end{figure}

A second option is normally preferable and is based on the constraint generation technique, widely used in integer programming. We build the set $\mathcal{R}$ as before, but for each convex polygon $R \in \mathcal{R}$ we add to $W$ a point inside $R$. Then, we run the solver again and repeat until a valid solution is found (or one with a very few uncovered regions). It is perhaps surprising how few iterations are normally needed, as shown in Figure~\ref{f:vwit}.

\subparagraph*{Set Cover Solver}

A simple and often efficient way to solve a set cover problem $(W,\CC)$ is to model the problem as \emph{integer programming (IP)} and then use the CPLEX solver~\cite{cplex}. Each set in $\CC$ becomes a binary variable and each witness point $w \in W$ becomes a constraint forcing the sum of the sets that contain $w$ to be at least $1$. The object is to minimize the sum of the variables. As discussed in the next section, this approach can optimally solve fairly large problems in seconds and give good approximation guarantees to some large problems. However, for some other large problems the solution found is extremely bad (sometimes worse than a greedy algorithm).

Another solver we use is based on \emph{simulated annealing}. This solver is a good fallback when the IP solver times out with a large gap between the lower bound and the solution found. The solver starts from a greedy solution, obtained by adding to $\SS$ the convex polygon that covers the most uncovered witnesses at each step, breaking ties randomly. Anytime, if a previously added convex polygon in $\SS$ becomes unnecessary, we remove it from $\SS$. At each step, we remove $3$ random convex polygons from $\SS$ and use the same greedy approach to make the solution cover all $W$.

A larger solution is accepted with a certain probability that depends on the size difference and decreases as we advance in the annealing procedure. To control this probability we set a \emph{temperature} $t=100/i$ for iteration number $i$. Let $d$ be the size of the previous solution minus the size of the new solution, all divided by the size of the previous solution. The probability that we accept the new solution is $e^{100d/t}$. This simple procedure normally produces solutions that are close to the IP solutions, and may produce much better solutions for some hard instances. However, it is much slower than CPLEX when solving easy instances, since we lack a mechanism to detect that an optimal solution has already been found.

\section{Results} \label{s:results}

We now discuss the quality of the solutions obtained for each technique. Our C++ code uses CGAL~\cite{cgal} and CPLEX~\cite{cplex} and is run on Fedora Linux on a Dell Precision 7560 laptop with an Intel Core i7-11850H and 128GB of RAM. All times refer to a single core execution with scheduling coordinated by GNU Parallel~\cite{parallel}. Our plots for a solution $\SS$ use the \emph{relative solution size}, defined as $|\SS^*|/|\SS|$, where $\SS^*$ is the best solution submitted among all teams. This corresponds to the square root of the Challenge \emph{score} of $\SS$.

\begin{figure}[ht]
 \centering
 \includegraphics[scale=.42]{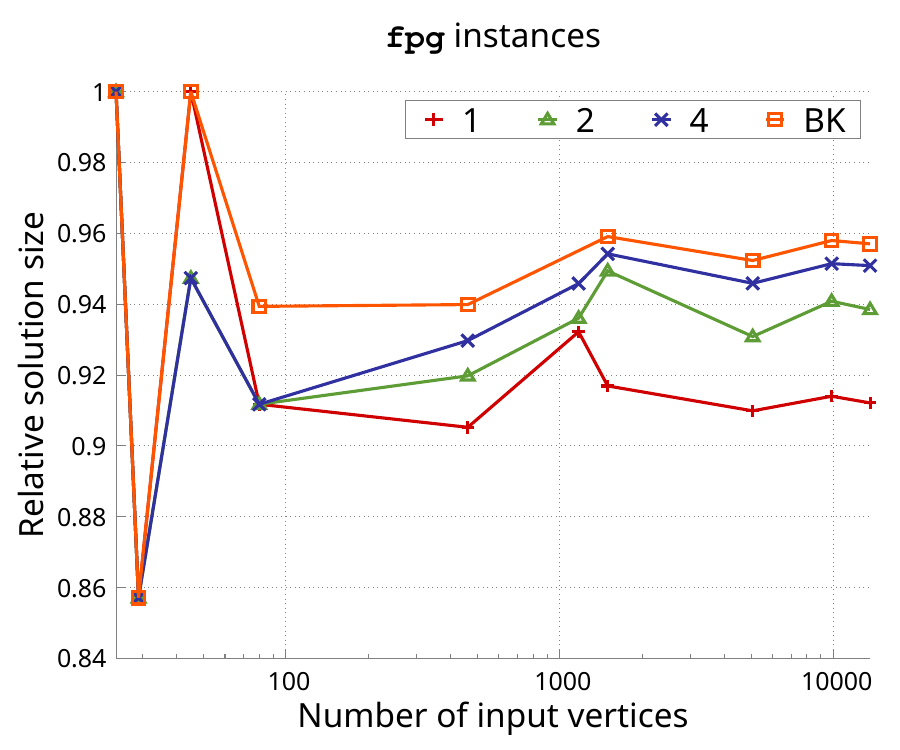} \hspace{2em}
 \includegraphics[scale=.42]{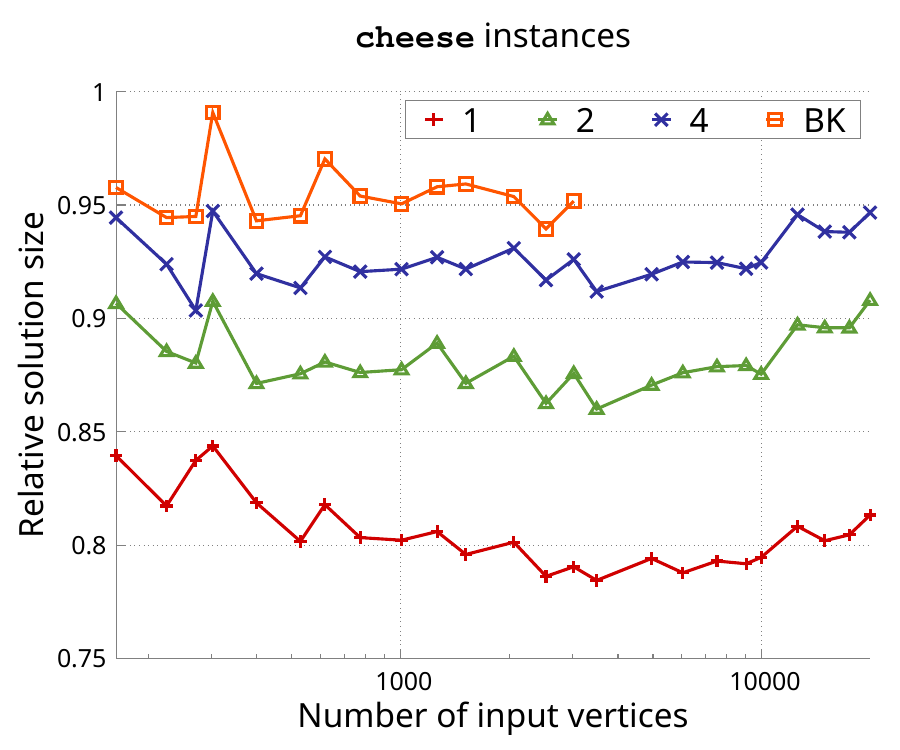}
 \caption{Solution sizes relative to the best Challenge solution. Data is based on a triangulation replicated $1$, $2$, or $4$ times and randomly bloated using $V$. Alternatively, we use Bron-Kerbosch (BK) to obtain all $V$-maximal polygons, when the running time is not too long. Afterwards, all collections are bloated again using $V \cup S_2(C)$ and solved using IP.}
 \label{f:grow}
\end{figure}

\begin{figure}[ht]
 \centering
 \includegraphics[scale=.42]{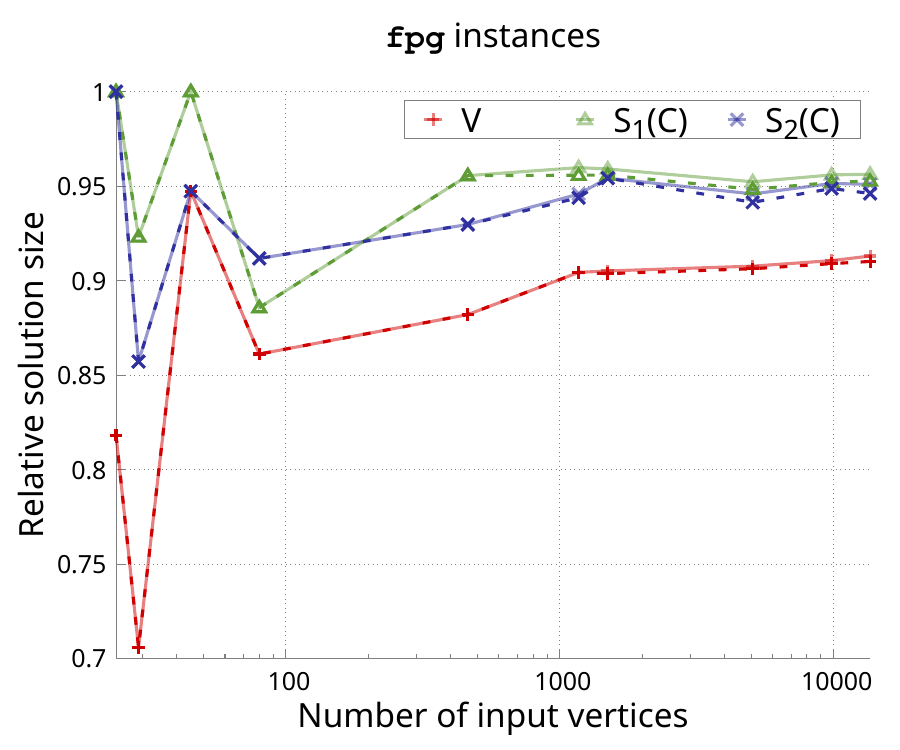} \hspace{2em}
 \includegraphics[scale=.42]{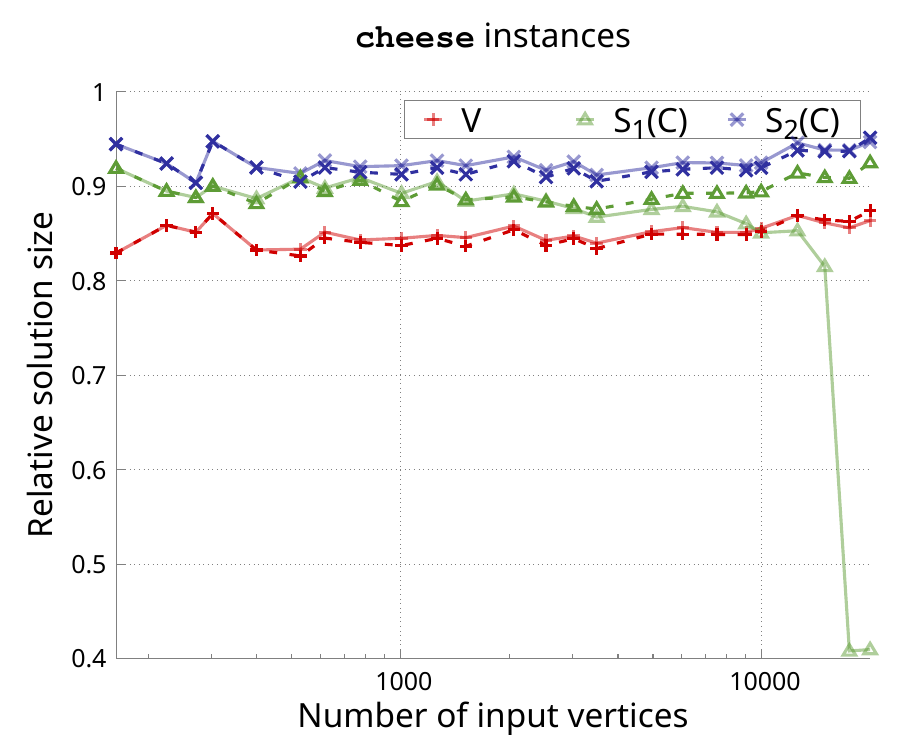}
 \caption{Solution sizes relative to the best Challenge solution. A solid line is used for the IP solver and a dashed line for simulated annealing. Data is based on a triangulation randomly bloated using $V$ (red) and then bloated again using $V \cup S_1(C)$ (green), or $V \cup S_2(C)$ (blue).}
 \label{f:sol}
\end{figure}

\begin{figure}[p]
 \centering
 \includegraphics[scale=.42]{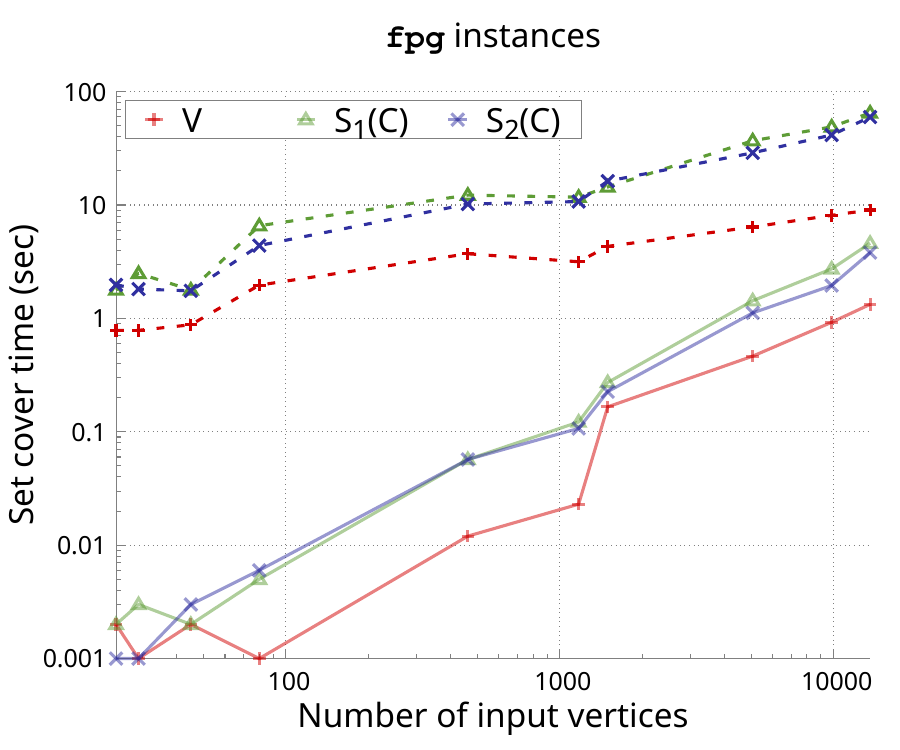} \hspace{2em}
 \includegraphics[scale=.42]{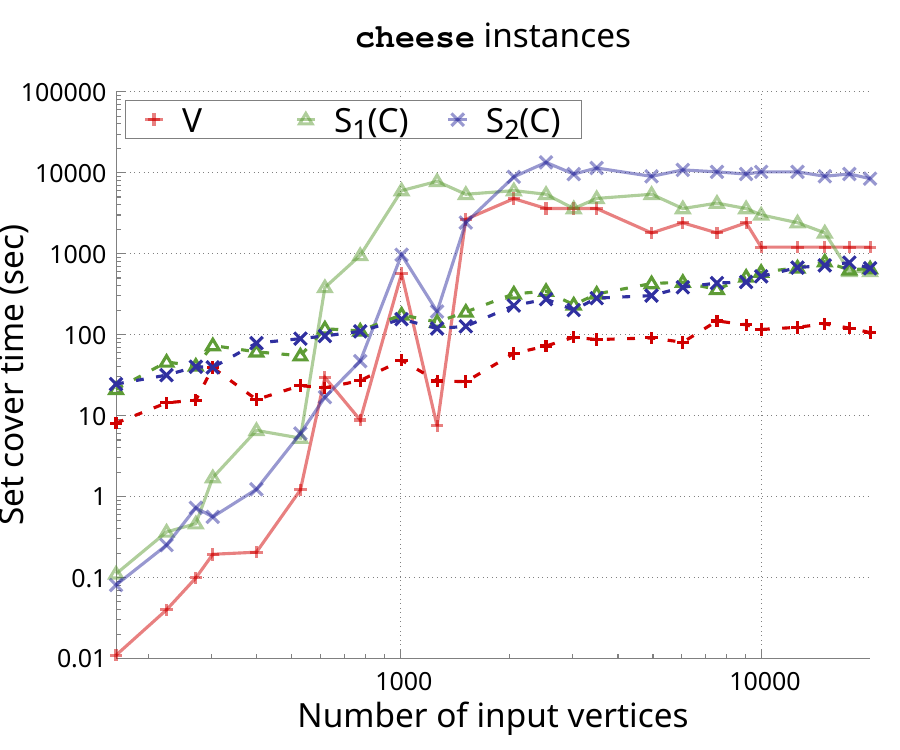}
 \caption{Running times in seconds of the set cover solvers used to find the solutions of Figure~\ref{f:sol}.}
 \label{f:time}
\end{figure}

Figure~\ref{f:grow} compares the different techniques to obtain $V$-maximal convex polygons before bloating them. As the figure shows, using $4$ replications of each constrained Delaunay triangle gives solutions that are almost as good as Bron-Kerbosch, but works on all instance sizes. Hence, we use this setting for Figures~\ref{f:sol} and~\ref{f:time}.

Figure~\ref{f:sol} shows the relative solution sizes using different bloating approaches and comparing the simulated annealing and the IP solvers. You can see that the simulated annealing solver is only slightly worse than IP for small instances, but better for large \texttt{cheese} instances. We limited the running time of IP to 10 minutes per iteration. The total running times of the solvers are compared in Figure~\ref{f:time}.

\begin{figure}[p]
 \centering
 \includegraphics[scale=.42]{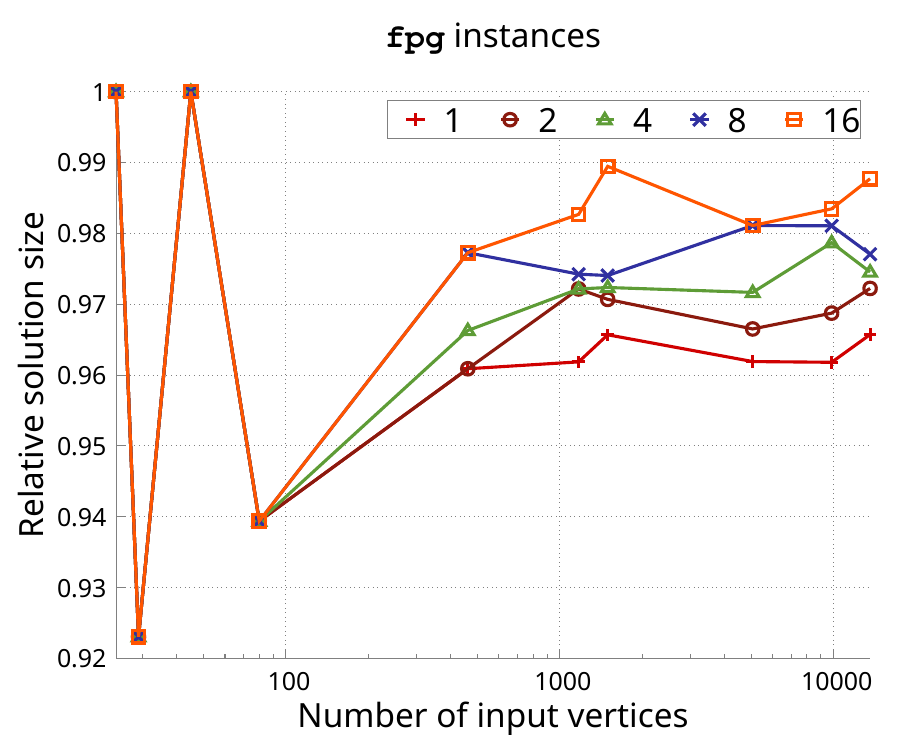} \hspace{2em}
 \includegraphics[scale=.42]{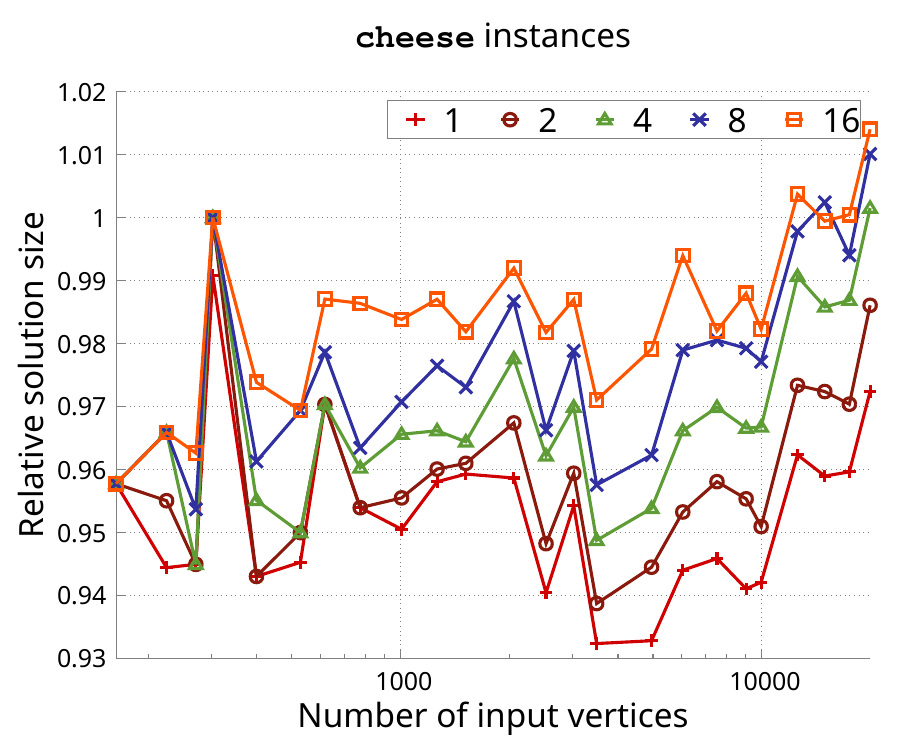}
 \caption{Relative solution sizes merging the $k$ best solutions for different values of $k$.}
 \label{f:merge}
\end{figure}

\begin{figure}[p]
 \centering
 \includegraphics[scale=.42]{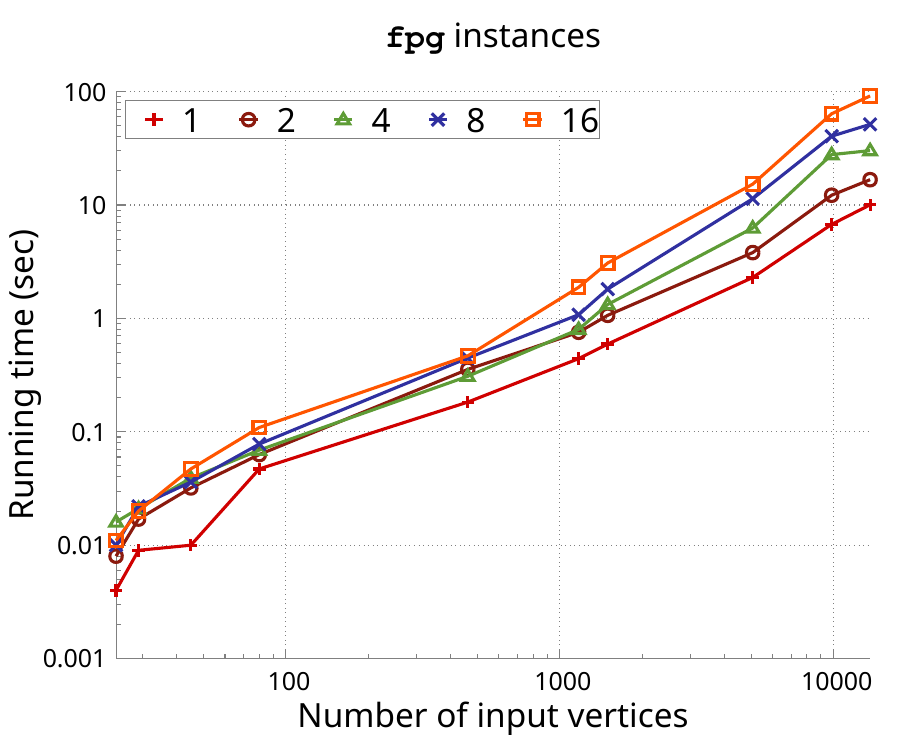} \hspace{2em}
 \includegraphics[scale=.42]{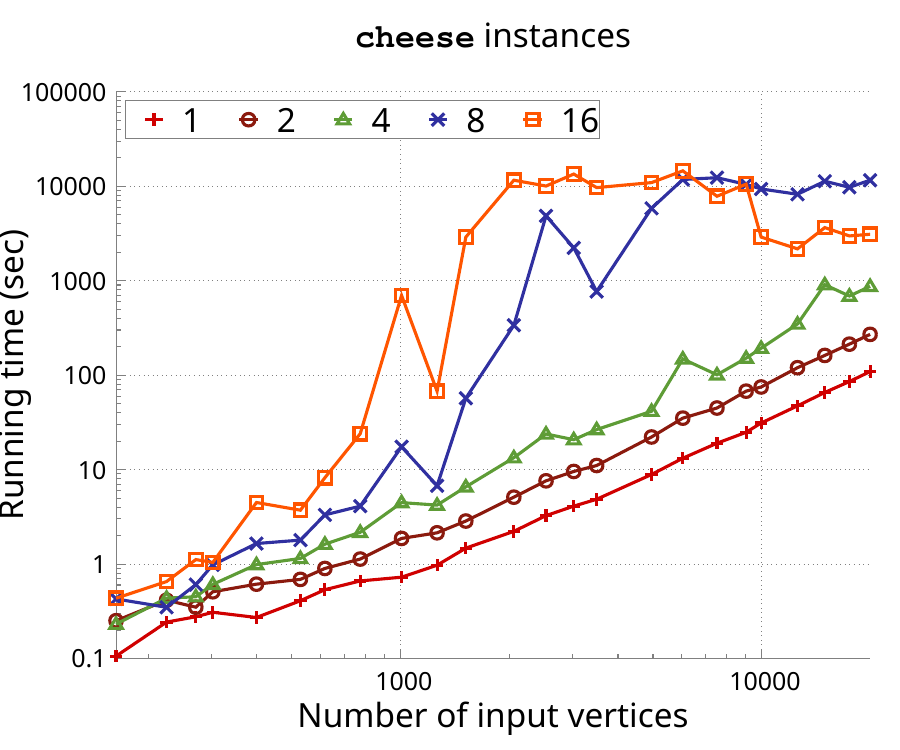}
 \caption{Running times in seconds to find the solutions of Figure~\ref{f:merge}.}
 \label{f:mergetime}
\end{figure}

A much better collection is obtained by using the union of several high quality solutions as a collection.
To produce the previous plots, we performed $3$ independent runs for each settings (showing the best result found). Figures~\ref{f:merge} and~\ref{f:mergetime} show the solution sizes  and times obtained by using the best $k$ solutions from these runs as the collection.

\section{Open Problems}

When solving phase $2$, we noticed that the number of iterations is often very small. This raises a few theoretical questions that we could not answer. First, is there a non-trivial upper bound on the number of iterations using vertex witnesses?

The number of witnesses needed to guarantee a correct solution is an intriguing question. Given a collection $\CC$ that covers the polygon $P$, how many witnesses are necessary? Notice that if the instance polygon is a $u \times u$ square and the $\CC$ consists of two tilings, one of $u$ rectangles of size $1 \times u$ and one of $u$ rectangles of size $u \times 1$, then $u^2$ witnesses are necessary. However, if we require the solution of the set cover problem to be optimal, then $O(u)$ witnesses suffice. More interestingly, what happens if $\CC$ is the collection made of all $V$-maximal polygons for the vertices $V$ of $P$? Is a subquadratic bound on the number of segments of $\CC$ possible?

The Bron-Kerbosch algorithm is a simple algorithm to enumerate all maximal cliques of a given graph. While it has good practical performance, it is not an efficient enumeration algorithm for several graph classes. Other algorithms with better theoretical guarantees are know for many graph classes. Are there provably efficient enumeration algorithms for the $V$-maximal convex polygons? Notice that, in the case of a polygon without holes, the problem is equivalent to enumerating the maximal cliques of the visibility graph.

\bibliography{references}

\end{document}